%% file: main.tex
\documentclass[11pt]{article}
\usepackage[margin=1in]{geometry}
\usepackage{authblk}

\usepackage[T1]{fontenc}
\usepackage[hidelinks]{hyperref}
\usepackage{cleveref}
\usepackage{enumitem}
\usepackage{amssymb}
\usepackage{upquote}

\newcommand{\N}{\mathbb{N}}

\usepackage{needspace}

\input{isabelle_listings}

\crefname{lstlisting}{listing}{listings}
\Crefname{lstlisting}{Listing}{Listings}

\usepackage{orcidlink}

\begin{document}

\title{A First Introduction to Isabelle/ML Metaprogramming\\[1ex]\large Automatic Estimation of Polynomial Degrees}
\author[1]{Jonas Bayer\,\orcidlink{0009-0006-2500-3146}\thanks{\texttt{jcb234@cam.ac.uk}}}
\author[2]{Anna Danilkin}
\author[3]{Marco David\,\orcidlink{0000-0001-5673-5389}}
\author[3]{Annie Yao}

\affil[1]{University of Cambridge, Cambridge, UK}
\affil[2]{École Normale Supérieure, Paris, France}
\affil[3]{University of California, Berkeley, USA}
\date{}

\maketitle

\begin{abstract}
  This article offers an introduction to metaprogramming in Isabelle/HOL for beginners, based on a running example for working with multivariate polynomials. The example is motivated by our formalisation of universal Diophantine pairs~\cite{up-itp}. We describe the implementation of the \lstinline|poly_degree| command, which computes upper bounds on the total degrees of multivariate polynomials and automatically proves their correctness. The complete metaprogram handles a variety of special cases but herein we present a simplified version for the sake of exposition. We describe our development process and design decisions---our goal is to offer a small and self-contained tutorial on Isabelle/ML, for mathematicians who want to get started with metaprogramming.
\end{abstract}

\parindent 0pt\parskip 0.5ex

\section{Introduction}

In this article, we present the basics of metaprogramming with Isabelle/ML. This is a first introduction to the subject, which exemplifies how to programmatically construct terms and propositions, declare new constants and theorems, and automate proving basic theorems about them. The accompanying code can be found on Zenodo.\footnote{\url{https://doi.org/10.5281/zenodo.20418298}}

This tutorial-by-example is based on the metaprogramming infrastructure developed for our previous formalisation of complexity bounds on universal Diophantine pairs~\cite{up-itp}. We used Isabelle as a true \emph{assistant} in constructing the proof, developing much of the formalisation~\cite{up-afp} in parallel with our mathematical draft. A corollary of our main result, relying on 50 pages of number theory, is a stronger version of Hilbert's Tenth Problem. In this tutorial, however, we assume no prior knowledge of Diophantine equations or universal pairs, beyond a basic understanding of multivariate polynomials.

\subsection{Motivation}

A key challenge of our formalisation was defining around 30 complicated, nested multivariate polynomials, whose definitions in our manuscript span multiple pages. In order to work with these polynomials both as ordinary Isabelle functions (i.e. of type \lstinline|int $\Rightarrow$ int $\Rightarrow$ ... $\Rightarrow$ int|) and as abstract polynomial objects of type \mbox{\lstinline|int mpoly|} at scale we first developed a metaprogram called \lstinline|poly_extract|. It translates between these two types for all Isabelle terms that actually look like a Diophantine polynomial, and also proves the correctness of this translation upon evaluation.

At a later point, it became necessary to calculate the total degrees of these intricate multivariate polynomials. Recall that the total degree of a monomial is simply the sum of the exponents of all variables; the total degree of a polynomial can then be defined as the maximum total degree over its monomials. We set out to develop the \lstinline|poly_degree| command to automatically calculate the degrees of our myriad polynomials. Moreover, the command should also prove that the calculated degree was correct.

More precisely, the command \lstinline{poly_degree} calculates upper bounds for these total degrees. The reason is that computing exact degree values can be non-trivial when cancellation occurs. For example, the polynomial $x^\delta y - x y$ can have degree zero if $\delta=1$. Because precise values for the degree were not required for our intended application, we settled on calculating upper bounds.

\subsection{Multivariate Polynomials in Isabelle}
Our formalisation is based on the Archive of Formal Proofs (AFP) entry on Executable Multivariate Polynomials~\cite{polynomials-afp}. This entry defines a type \lstinline{'a mpoly} for polynomials with coefficients of type \lstinline|'a| using maps with finite support. However, we will not need to understand this abstract definition directly. Instead, we work with two basic constructors: the constant polynomials \lstinline|ConstP :: 'a $\Rightarrow$ 'a mpoly| and the single-variable monomials \lstinline|VarP :: nat $\Rightarrow$ 'a mpoly|. Since the \lstinline|mpoly| type instantiates the plus and times typeclasses, all polynomials can be built using just \lstinline|ConstP|, \lstinline|VarP|, and these operators.\footnote{The names \lstinline|ConstP| and \lstinline|VarP| used here are aliases for \lstinline|Const| and \lstinline|Var| in the original AFP entry. We do this to avoid confusion with the constructors of the ML type \lstinline|term| of the same name.} An example can be seen in \Cref{lst:example-polynomial}.
\isasnippet[caption={\label{lst:example-polynomial}Formalisation of the polynomial $P(x_0, x_1, x_2) = x_1^2 \cdot x_2 + x_0 + 3$}.]{polynomial-example}

\subsection{An Overview of \texttt{poly\_degree}}

We now provide a high-level description of the command \lstinline|poly_degree| that we will implement in this text. One can invoke it on any Isabelle constant of type \lstinline|int mpoly| as follows:

\begin{lstlisting}[language=isabelle]
poly_degree P
\end{lstlisting}

When invoked like this, the command creates a new constant \lstinline|P_degree| together with a theorem that states that the total degree of \lstinline|P| is bounded above by \lstinline|P_degree|. Note that here ``constant'' is to be understood in the sense of a constant in the logic. In particular, \lstinline|P_degree| does not have to be a numerical constant, but could also be an expression like $\delta + 1$ if $\delta$ is a fixed variable in the context where \lstinline|P| is defined. After invoking \lstinline|poly_degree|, both the constant \lstinline|P_degree| and the generated theorem \lstinline|P_degree_correct| can be used as if they had been declared using the \lstinline|definition| or \lstinline|lemma| commands. Furthermore, the generated theorem is proved automatically.

This may seem more complicated than necessary: one might compute an explicit bound and use it directly, without defining a separate constant. Although this approach works in many situations, it would have quickly led to a blowup in complexity in our project, making proofs infeasible. To illustrate this, consider polynomials $P$ and $Q$ with complicated degree expressions. When we define $R = P + Q$, we can organise the resulting complexity much better by introducing new constants and obtaining a simple theorem \lstinline|total_degree R $\leq$ max P_degree Q_degree|. Introducing constants that hide complicated degree expressions makes our proof development more modular.

The following steps are executed by the metaprogram behind \lstinline|poly_degree|:
\begin{enumerate}
  \item Look up the defining term of \lstinline|P| from the context (\Cref{subsec:get-poly-info}).
  \item Use this term to generate a term \lstinline|P_degree| representing a degree bound for \lstinline|P| (\Cref{subsec:gen-term}).
  \item Generate a correctness theorem which states that the total degree of \lstinline|P| is bounded by \lstinline|P_degree| (\Cref{subsec:gen-statement}).
  \item Automatically prove the correctness theorem (\Cref{sec:proving}).
\end{enumerate}

After invoking the
command on \lstinline|P|, one can use the theorem \lstinline|P_degree_correct| as if it had been stated as follows:

\begin{lstlisting}[language=isabelle]
lemma P_degree_correct: "total_degree P $\leq$ P_degree"
\end{lstlisting}

\paragraph{Remark.} This tutorial presents a version of the \lstinline|poly_degree| command that is simplified, compared to the version published in the AFP, in the interest of a pedagogical exposition. The full version has special support for dealing with polynomials substituted into other polynomials. An effect similar to that of the simplified command presented here can also be obtained through the use of \lstinline|schematic_goal|, without a dedicated metaprogram. A schematic goal like \lstinline|total_degree P $\leq$ ?d| allows introducing a schematic variable \lstinline|?d| (i.e., a placeholder for the degree) and then proving the resulting statement using the simplification rules discussed in \Cref{sec:proving-bounds}. This calculates a valid bound in the process.

\subsection{Related Work and Our Contributions}
The canonical and always up-to-date reference for metaprogramming in Isabelle is the implementation manual by Wenzel~\cite{isabelle-implementation}. Moreover, Urban's Cookbook~\cite{urban-cookbook} aims at providing a more gentle tutorial in Isabelle metaprogramming. Note that this is not part of the official Isabelle documentation so might not be entirely in sync with the latest version of Isabelle. 

Both texts span more than 150 pages and cover the topic of metaprogramming in great detail. In contrast, this article aims to give a short and succinct introduction to Isabelle/ML, specifically directed at mathematicians who want to get started with metaprogramming in Isabelle. We focus on one single worked example, bounding polynomial degrees, to cover the following Isabelle/ML interfaces: extracting definitional equations from the context, constructing terms and propositions programmatically, and introducing new constants. We also illustrate how ML tactics can mirror apply-style proofs. Where appropriate, we provide pointers to the implementation manual and source code for further reading.

We observe that in the Lean ecosystem developing domain-specific tactics, which very effectively solve a narrowly defined set of goals, has been a successful strategy for aiding the formalisation of advanced mathematics. For example, there are specialised tactics that target continuity, measurability or goals in category theory.\footnote{\url{https://leanprover-community.github.io/mathlib4_docs/tactics.html}} We hope that this text will reduce the entry barrier for those who want to explore a similar approach in Isabelle. 

\paragraph{Limitations.} We only intend this text to be a first stepping stone that makes it easier for learners to get started with Isabelle/ML. We emphasize that writing maintainable, high quality code as would be necessary for, e.g., contributing to Isabelle itself, is outside of our scope and requires significant extra learning and experience.

\section{An Isabelle/ML Crash Course}
In this tutorial, we want to interact with Isabelle programmatically: transform terms, create new definitions, and prove theorems. All of this is possible through Isabelle's ML interface, so we will start with an Isabelle/ML crash course.

In what follows, we assume that the reader is familiar with Isabelle/HOL. Many concepts that we encounter, e.g., pattern matching, should be easy to understand for this audience without a detailed introduction to ML. We explain ML syntax where it is not immediately obvious and refer the reader to Paulson's book~\cite{paulson-ml} for an in-depth introduction to programming in ML. 

\subsection{Basic Interaction with the ML Interface}

In this text, we always use ML blocks to write Isabelle/ML inside an Isabelle theory. ML blocks can be placed anywhere outside a proof; an example is shown in the listing below.\footnote{In the following listings, we often omit the opening (\lstinline[language=isabelle]{ML\\<open>}) and closing (\lstinline[language=isabelle]{\\<close>}) of the block itself, in which case the language of the listing is marked as ``ML.''}

\isasnippet{happy-birthday}

Placing the cursor on this code inside an Isabelle theory will show a congratulatory message in the output pane. Note that it is also possible to write ML code in separate files and then load them into Isabelle using \lstinline|ML_file\<open>...\<close>|.

Before we give a brief overview of the main low-level concepts used in metaprogramming, we first recall the following important Isabelle concepts.

\begin{itemize}
  \item \textbf{Inner and Outer Syntax}. In Isabelle, there is a constant interplay between an outer and inner layer of the language: the outer level includes commands like \lstinline|definition|, \lstinline|lemma| etc. as well as structured proofs in Isabelle's proof language Isar. The inner level is essentially everything that comes inside quotation marks---the terms and types of the logic about which one writes proofs.
  \item \textbf{Pure Logic and Isabelle/HOL}. Recall that Isabelle is a framework for implementing various logics, with HOL being a specific example. One encounters two ``levels'' of implication $\Longrightarrow$ and $\longrightarrow$, corresponding to Pure implication and HOL implication, and these two separate layers will also be present on the ML side.
  \item \textbf{Notation, Short and Long Names}. In order not to clutter up one single global namespace, names are managed in a hierarchical fashion. For example, group multiplication in HOL would be \lstinline|Groups.times_class.times g h| if one were to use the long, fully qualified name. In practice, as long as there is no ambiguity, one can also write \lstinline|times g h|. Moreover, with user defined mixfix syntax, this shortens further to \lstinline|g * h|. Crucially, all these syntax expressions denote the same logical entity.
\end{itemize}

\subsection{Isabelle's Type and Term Datatypes}
Some of the most important ML datatypes encountered in Isabelle metaprogramming are \lstinline|typ| and \lstinline|term|, representing the types and terms of the logic. Every piece of syntax inside quotes is ultimately parsed and type-checked into a value of one of these datatypes.

The type \lstinline|typ| has three constructors. The constructors \lstinline|TFree| and \lstinline|TVar| are used to represent type variables and schematic type variables, and they are equipped with type class information. The constructor \lstinline|Type| is used for constructing new types from existing types. It takes a list of types and the name of the type constructor as arguments. This is how function types and inductive types are represented.

In practice, one often creates values of type \lstinline|typ| by using the \lstinline|@{typ ...}| antiquotation. For example, we will write \lstinline|@{typ int}| to obtain the ML value representing the Isabelle/HOL type \lstinline|int|. We show more examples of antiquotations at the end of this section. We refer to the Isabelle implementation reference~\cite[Section 2.1]{isabelle-implementation} for all details on \lstinline|typ|.

The ML datatype \lstinline|term| is used to internally represent the terms of Isabelle's logic. At a high level, \lstinline|term| has a tree structure that captures variables, constants, binders, and function application. Its definition is as follows:

\begin{mllisting}[mathescape,caption={Definition of Isabelle's term datatype.}]
  datatype term =
  Const of string * typ |
  Free of string * typ |
  Var of indexname * typ |
  Bound of int |
  Abs of string * typ * term |
  $\mbox{\textdollar}$ of term * term
\end{mllisting}

Each constructor has a specific role, which we explain below for completeness. Note, however, that we only need to deal with constants and function application for the implementation of \lstinline|poly_degree|.

\begin{itemize}
  \item \lstinline|Const (c, T)| represents a globally defined constant named \lstinline|c| of type \lstinline|T| (e.g. the constant \lstinline|0| or \lstinline|plus|). In their term representation, constant names are fully qualified: the name \lstinline|c| for \lstinline|plus| would be \lstinline|Groups.plus_class.plus|.
  \item \lstinline|Free (x, T)| represents a free variable named \lstinline|x| of type \lstinline|T|. For example, the ML value of the term \lstinline|x::int|, i.e., an integer variable with name $x$, would be \lstinline|Free ("x", "int")|.
  \item \lstinline|Var ((x, i), T)| represents a schematic variable of type \lstinline|T|, written \lstinline|?x.i| in Isabelle. The pair \lstinline|(x, i)| is an \lstinline|indexname|: the index \lstinline|i| distinguishes variables with the same base name.
  \item \lstinline|Bound n| represents a bound variable using de Bruijn indices. That is, the index $n$ denotes to which lambda binder the variable belongs. We explain de Bruijn indices in more detail below.
  \item \lstinline|Abs (x, T, t)| represents a lambda abstraction \lstinline|$\lambda$x::T. t|. The name \lstinline|x| is only for readability; binding is determined by the indices of \lstinline|Bound| that appear in the term \lstinline|t|.
  \item \lstinline[mathescape=false]|f $ u| represents function application. For instance, the term \lstinline|f x y| is encoded as \lstinline[mathescape=false]|f $ x $ y|.
\end{itemize}

To give a simple example, the ML value for the term \lstinline|1 + (x :: nat)| looks as follows:

\begin{lstlisting}[language=ML,caption={ML representation of $1 + (x : \N)$.},label={lst:term-one-plus-x}]
Const ("Groups.plus_class.plus", "nat \<Rightarrow> nat \<Rightarrow> nat") 
$ Const ("Groups.one_class.one", "nat") $ Free ("x", "nat")
\end{lstlisting}

To understand this representation, it is better to think of the term as \lstinline|plus 1 x| instead of using the infix (\lstinline|+|) operator. Just like with functions in HOL, function application (\lstinline[mathescape=false]|$|) is curried.

\subsection{Binders and De Bruijn Indices}
A naive way of implementing bound variables would be to simply assign them a name which could be referenced in the binder. However, this creates a lot of complications when dealing with multiple variables: one has to make sure that the names are unique. For example, consider the statement $\forall x \in \N.\; f\; x = 0$ where $f$ is a constant (of the logic) defined as $\lambda x.\; x + 1$. If one were to unfold the definition of $f$ in the statement, there would be ambiguity about where each $x$ is bound.

An elegant solution developed to remedy this problem is to keep the bound variables entirely nameless~\cite{deBruijn1972}. Instead, they simply receive an index that says how many levels of binders one has to go up to arrive at the binder corresponding to this specific variable. For example, $\lambda x : \N.\; \lambda y : \N.\; f \; x \; y$ then becomes the following:
\begin{lstlisting}[language=ML]
Abs ("x", "nat", Abs ("y", "nat", 
  Const ("f", "nat \<Rightarrow> nat \<Rightarrow> nat") $ Bound 1 $ Bound 0))
\end{lstlisting}
If we swapped $0$ and $1$ in this code, the corresponding term would be $\lambda x.\; \lambda y.\; f \; y \; x$. The indices of the bound variables are also called \emph{de Bruijn indices}.

Note that in Isabelle, the only actual binder is lambda abstraction \lstinline|Abs|. All other binders, like $\bigwedge, \forall, \exists$ etc., are represented by constants that take lambda terms as arguments. For example, consider the (false) statement $\forall x.\; x$ where $x$ is of type \lstinline|bool|. It would be represented as follows:

\begin{lstlisting}[language=ML]
Const ("HOL.All", "(bool \<Rightarrow> bool) \<Rightarrow> bool") $ Abs ("x", "bool", Bound 0)
\end{lstlisting}

Here, the constant \lstinline|HOL.All| has the type \lstinline|(bool $\Rightarrow$ bool) $\Rightarrow$ bool|: it expects a predicate on booleans and returns a boolean. Hence we pass it the predicate $\lambda x.\; x$ of type \lstinline|bool $\Rightarrow$ bool|, represented by \lstinline|Abs ("x", "bool", Bound 0)|.

\subsection{Antiquotations}\label{subsec:antiquotations}
Writing terms naively using term constructors like \lstinline|Const| can be unsafe: if someone were to rename one of the constants this could lead to unpredictable behaviour of the ML code. This is a general problem that doesn't just exist for terms but also theorem identifiers, types, and more.

Isabelle provides antiquotations as a solution for this issue. The implementation manual describes: ``A regular antiquotation \lstinline|@{name args}| processes its arguments by the usual means of the Isar source language, and produces corresponding ML source text''~\cite[Section~0.2.3]{isabelle-implementation}.

For example, to create the term shown in \Cref{lst:term-one-plus-x}, one can use the \lstinline|@{term ...}| antiquotation:

\mlsnippet{one-plus-x-antiquotation}

A good way to learn the internal representation of terms is to inspect examples and view their ML encoding in the output pane using the \lstinline|term|-antiquotation.

In addition to terms and types, antiquotations are also useful for referencing theorems and for debugging output. The antiquotation \lstinline|@{thm name}| retrieves a checked theorem value of type \lstinline|thm|. For example, \lstinline|@{thm le_trans}| retrieves the transitivity of less than or equal. Another commonly used antiquotation is \hbox{\lstinline|@{make_string}|}: it turns ML values into readable strings, which is convenient for debugging purposes. The antiquotation \lstinline|@{print}| can be used to display ML values in the output pane. We will see examples of concrete use cases for all these antiquotations in the following sections.

\section{Computations on Terms}
Before we present the actual \lstinline|poly_degree| command, we start by implementing a simpler toy command: \lstinline|poly_degree0| calculates an upper bound for the degree of a given polynomial and writes the result to the output pane.

The subsequent listings assume that the theory \lstinline|Setup| from the code accompanying this text has been imported. It contains the necessary AFP imports for multivariate polynomials as well as auxiliary aliases (\lstinline|ConstP|, \lstinline|VarP|) for this tutorial.

\subsection{Computing Numerical Degrees}
The heart of \lstinline|poly_degree0| is a function that takes an Isabelle term representing a polynomial and returns a bound on its degree as an ML integer.

\mlsnippet[caption={Computing the degree for a given polynomial term.}]{compute-degree-function}

Most of the work is done by pattern matching directly on the term. We use the \lstinline|@{Const ...}| antiquotation to match fully applied constants together with their arguments in one step. If the given term~\lstinline|t| is a polynomial, it should match one of the constructors \lstinline|ConstP| (constant polynomial), \lstinline|VarP|, or the operations \lstinline|plus| and \lstinline|times|.

In line~3, we can see the pattern for matching constant polynomials. The \lstinline|@{Const ...}| pattern is checked by Isabelle and avoids having to manually work with fully qualified constant names. The wildcard (\lstinline|_|) is used in \lstinline|"ConstP _ for _"| to ignore both the type argument and the concrete coefficient, since neither affects the degree.

The \lstinline|VarP| case is analogous to the \lstinline|ConstP| case. For sums, we recursively compute the degrees of both subterms and take the maximum; for products, we add the degrees.

If we encounter a term which does not correspond to one of the previous cases we simply raise an error. Note that \lstinline|^| is string concatenation in ML and we use \lstinline|@{make_string}| to turn the given term into a printable string.

\subsection{Testing the Computation Function}
To test our function we can use the \lstinline|@{term ...}| antiquotation:

\mlsnippet{test-compute-degree}

We can also apply the function to terms extracted from definitions in our Isabelle theory. Assuming that a polynomial \lstinline|P| was defined as in \Cref{lst:example-polynomial}, the following ML code will calculate its degree.

\mlsnippet{compute-degree-example}

In the first line, we retrieve the theorem using an antiquotation. We can use \lstinline|Thm.prop_of| to extract the underlying term, which should look like \lstinline|P $\equiv$ rhs|, where \lstinline|rhs| is the defining term for \lstinline|P|. By using \lstinline|Logic.dest_equals|, we can then obtain an ML tuple \lstinline|(P, rhs)| from it. Finally, we apply \lstinline|compute_degree| to the term \lstinline|rhs| and print it to the output pane using \lstinline|@{print}|.

Note that the first three lines of the above listing can be written more succinctly using the pipe operator \lstinline{|>} as follows. Further operators like this can be found in \texttt{\~{}\~{}/src/Pure/General/basics.ML}.

\mlsnippet{compute-degree-succinct}

\subsection{Defining Custom Commands}

While working directly in ML blocks is useful for development and testing, regular Isabelle users typically want a more convenient interface. To this end, it is possible to define custom Isabelle commands, extending the outer syntax of Isabelle. We now implement a command \lstinline|poly_degree0| which, if invoked as in the listing below, prints a bound on the total degree of the given polynomial $3 + x_0\cdot x_1$ to the output pane. 

\isasnippet{run-poly-degree0}

Note that to implement a purely diagnostic command one would normally use \lstinline|Toplevel.keep| \cite[Section~9.1.2]{isabelle-implementation}. However, since we want to add non-diagnostic capabilities to the command in the following section, we already set \lstinline|poly_degree0| up in a way that will allow us to do this. 

First, we need to declare the command keyword in the header of our theory file to make Isabelle's parser recognise it as such. There are different classifications for keywords, e.g. corresponding to whether the keyword is allowed to create proof obligations or not. We classify the \lstinline|poly_degree0| keyword as \lstinline|thy_defn| so we can later give it the capability to make new definitions.

\isasnippet[label={lst:theory-header},caption={Isabelle theory header with keyword declaration.}]{theory-header}

We then use \lstinline|Outer_Syntax.local_theory| to provide the actual command implementation. Its first two arguments are the command keyword itself and a documentation string. The implementation of the command is kept in a separate function \lstinline|poly_degree0_cmd|.\footnote{Note that it is generally best to separate parsing from the actual command logic; this way, the logic can also be called from ML directly. We will illustrate this when presenting the finished command \lstinline|poly_degree|.}

\mlsnippet[caption={The implementation of \lstinline|poly_degree0|.}]{poly-degree0-command}

The syntax \lstinline|Parse.term >> poly_degree0_cmd| instructs Isabelle that the command expects a single argument, which should be a term. We omit the details of outer syntax parsing here.\footnote{cf. \texttt{\~{}\~{}/src/Pure/Isar/parse.ML} and \texttt{\~{}\~{}/src/Pure/General/scan.ML}}

Note the arguments of \lstinline|poly_degree0_cmd|: it takes a string representation of the given term (the inner syntax still needs to be parsed) and an argument of type \lstinline|local_theory|. The latter contains information about the \emph{context}, which we explain in the following section. If we wanted to, e.g., make a new definition, we could do this by returning a modified local theory. For the time being, we simply return \lstinline|lthy| without making changes.

It is often handy to use let bindings when writing more complex ML functions. We can define auxiliary values step by step in the binding (the \lstinline|let ... in| part) and then refer to them in the body (\lstinline|in ... end|). In the case of \hbox{\lstinline|poly_degree0_cmd|,} we implement the logic for printing the degree in the binding and return the unmodified local theory in the let body.

In the binding, we first use the function \lstinline|Syntax.read_term| to turn the given string into an actual term object and infer its type. This also takes the local theory as an argument which is needed for processing the term. We can then calculate the degree with \lstinline|compute_degree| and print it using \lstinline|writeln|.

We now have a toy command \lstinline|poly_degree0| that can print a bound for the degree of the given polynomial term. The following section explains how to extend it to make new declarations.

\section{Declaring Constants and Theorems}\label{sec:declaring}
So far we have only extracted information from the Isabelle theory. In this section we will learn how to interact with it, adding new definitions and theorems.

\subsection{Context and Local Theory}
An important concept in Isabelle/ML is that of context. The context has information about which constants have been defined and which theorems have been proven. Naturally, the context changes as one progresses through a theory---one cannot access a constant or prove something about it before it has been declared.

A lot of internal operations of the theorem prover need to take the context as input. For example, both parsing and type checking require the context so that, e.g., user-defined mixfix syntax can be processed.

In Isabelle, one generally deals with context through the two ML types \lstinline|Proof.context| and \lstinline|local_theory|. Both carry the same information content but are used in different situations. Generally, \lstinline|local_theory| supports all operations that \lstinline|Proof.context| supports. In addition, it provides interfaces that are not available with \lstinline|Proof.context|, such as making a new definition or initialising a new proof. Through this setup, Isabelle ensures that prove-mode operations, operating on \lstinline|Proof.context|, cannot modify the environment in unintended ways. More details can be found in the implementation manual~\cite[Chapter~8]{isabelle-implementation}.

\subsection{High-level Overview of the \lstinline|poly_degree| Metaprogram}
We continue with a high-level overview of the implementation of the \lstinline|poly_degree| command. Again, it is necessary to declare a keyword \lstinline|"poly_degree" :: thy_defn| in the theory header as in \Cref{lst:theory-header}. The following listing shows the implementation of the command. It makes use of various utility functions, specifically developed for this command, which we explain in what follows.

\needspace{2\baselineskip}

\mlsnippet[caption={Definition of the \lstinline|poly_degree| command.},label={lst:poly-degree}]{poly-degree-cmd}

In line~4, the command adds a new constant to the context, which represents the degree bound. Following this, in line~5, the statement of the correctness theorem is generated, which is then proven in line~6. Finally, in line~7, the newly proven theorem is added to the context.

Note that only line~4 and line~7 modify the context: compared to \lstinline|lthy|, \lstinline[breaklines=false]|lthy'| contains an extra constant declaration and only \lstinline|lthy''| has the correctness theorem. All other operations do not affect the context. Note that the metaprogramming framework of Isabelle allows proving a theorem without making it available for future use; the function \lstinline|note_correctness_thm| is essential for the command to work.

Note that we create the correct logical objects from the user-provided identifier in a separate function \lstinline|get_poly_info|. This parsing is executed via \lstinline|poly_degree_cmd| when the command is invoked. The separate definition of \lstinline|poly_degree| makes it possible to also execute the same logic directly from ML.

\paragraph{Remark.} A robust implementation of this command should involve additional functionality to manage naming more carefully. For example, it should check that the generated names do not already exist or let the user provide a custom name. We refer to Section 1.2 in the implementation manual for more details on managing names~\cite{isabelle-implementation}.

\bigskip

Let's now look at each function in detail, starting with retrieving information from the current context. Henceforth, for simplicity of explanation, we will assume that we have defined a generic polynomial \lstinline|P :: int mpoly| to which we apply this command.

\subsection{Extracting Auxiliary Information}\label{subsec:get-poly-info}
\mlsnippet[caption={Extracting the necessary information from the context.}]{get-poly-info}

The \lstinline|get_poly_info| function extracts all the information we need in subsequent steps from the definition. Given the name of a polynomial, we retrieve its definitional theorem using \lstinline|Proof_Context.get_thm|. If our polynomial is named \lstinline|P|, we assume that the corresponding definition is stored under the name \lstinline|P_def|, which is what \lstinline|Thm.def_name| produces. Once we have the definitional theorem, we can extract both sides of the equation: the left-hand side gives us a term that is just the constant \lstinline|P| itself, and the right-hand side gives us its defining expression. We also extract the base name (without any module qualifiers), which we will use when naming the constant introduced in the next step.

\subsection{Generating the Term Representing the Degree Bound}\label{subsec:gen-term}

The \lstinline|define_degree_const| function adds a new constant to the theory.

\mlsnippet[caption={Declaring a constant representing the degree bound.}]{define-degree-const}

We compute the degree term in a separate function \lstinline|compute_degree_term| (see below) and create a binding. A binding is an abstract description of a name that is about to be declared. It is used to manage namespaces and also contains information about the current source code position (this is used to make Ctrl/Cmd+Click work behind the scenes).

For a polynomial \lstinline|P| we want the new constant to be called \lstinline|P_degree|. The appropriate string is obtained using the ML concatenation operator (\lstinline|^|) and we turn it into a binding using \lstinline|Binding.name|. Now, if the command is used in a theory \lstinline|MyTheory|, then Isabelle will automatically qualify the name to \hbox{\lstinline|MyTheory.P_degree|.}

We use \lstinline|Local_Theory.define| to add the definition to the context. This works analogously to using the \lstinline|definition| command in Isabelle. Because the signature is somewhat involved, let's unpack it step-by-step:

\begin{lstlisting}
  Local_Theory.define: (binding * mixfix) * (Attrib.binding * term) -> local_theory -> (term * (string * thm)) * local_theory
\end{lstlisting}

The function takes three arguments. The first is a pair \lstinline|(binding, mixfix)| where \lstinline|binding| specifies the name for the new constant and \lstinline|mixfix| provides optional syntax annotations. In our example, we use \lstinline|NoSyn| to indicate that no special syntax should be introduced.

The second argument \lstinline|(Attrib.binding, term)| controls the name of the definitional theorem and the defining term. Here, \lstinline|Attrib.binding| is itself a pair consisting of a binding for the theorem name and a list of attributes (such as \lstinline|[simp]|). In our case, we use \lstinline|Thm.def_binding binding| to generate the standard definitional theorem name \lstinline|P_degree_def|, and we pass an empty attribute list \lstinline|[]|. The second component is the defining term, the right-hand side of the definition, which we computed using \lstinline|compute_degree_term|.

The third argument is the current local theory context.

The function returns a pair, where the first component identifies the newly created constant and the second component is the updated local theory context. More specifically, the first component of type \lstinline|term * (string * thm)| is itself a pair containing the newly created constant as a term and yet another pair of its fully qualified name and its definitional theorem. In our implementation, we extract the constant as \lstinline|degree_const| and the definitional theorem as \lstinline|degree_def|, ignoring the fully qualified name. These are the relevant values that we will use in the following steps.

It remains to explain the \lstinline|compute_degree_term| function which computes the actual bound.

\mlsnippet[caption={Computing the degree bound term.}]{compute-degree-term}

The function \lstinline|compute_degree_term| is similar to our earlier \lstinline|compute_degree| function, but we additionally add support for powers. The main difference is that instead of returning an ML integer, it constructs an appropriate Isabelle term of type \lstinline|nat|.

For constant polynomials, we return the term \lstinline|0::nat| and for variables, we return \lstinline|1::nat|, making use of term antiquotations. For sums of polynomials, we take the maximum of the degrees of the summands; for products, we add them; for powers, we multiply with the exponent. To build the appropriate term for \lstinline|plus|, we use the antiquotation \lstinline|@{const max(nat)}| which creates a constant term. Note that \lstinline|max| is polymorphic so we pass the type \lstinline|nat| in parentheses to specialise it to natural numbers.

\subsection{Generating the Degree Bound Statement}\label{subsec:gen-statement}

With the degree term in hand, we can now construct the statement that it is actually a bound on the total degree of the given polynomial.

\mlsnippet[caption={Constructing the degree bound statement.}]{mk-correctness-thm}

The \lstinline|mk_correctness_thm| function takes the polynomial constant and the degree term and builds the proposition \lstinline|total_degree poly_term| $\leq$ \lstinline|degree_term|. We first define the left-hand side and right-hand side terms of this theorem separately.

For the left-hand side, we need to use \lstinline|total_degree|, which normally has type \lstinline|'a::zero mpoly $\Rightarrow$ nat|. Since we focus only on polynomials over the integers here, we specialise \lstinline|total_degree| to integer polynomials in the antiquotation by giving the type \lstinline|int| in parentheses. The right-hand side is simply the provided degree term.

Finally, we can use \lstinline|HOLogic.mk_binrel| as a convenient way to create the inequality term. Note that this yields a HOL term that type-checks as \lstinline|bool|. However, when starting a new proof, Isabelle's kernel always expects a term of type \lstinline|prop|. We therefore wrap the term with \lstinline|HOLogic.mk_Trueprop|, which turns HOL booleans into Isabelle/Pure propositions.

The next function that gets called in \Cref{lst:poly-degree} is \lstinline|prove_correctness|, in charge of proving the theorem we just generated. It takes the proposition generated above and returns a value of type \lstinline|thm| if the proof succeeds. We explain it in \Cref{sec:proving} on proving in Isabelle/ML.

\subsection{Adding the Proven Theorem to the Local Theory}
Having used \lstinline|prove_correctness| we obtain a value of type \lstinline|thm|. The last step of the \lstinline|poly_degree| metaprogram is to add this theorem to the context so it will be accessible to the user. This is achieved through the following function.

\mlsnippet[caption={Adding the degree bound statement to the local theory.}]{note-correctness-thm}

The \lstinline|note_correctness_thm| function adds the theorem to the context under the name \lstinline|P_degree_correct|. This is similar to writing a \lstinline|note| command in Isar---the theorem becomes part of the theory and can be referenced in subsequent proofs.

The interesting part is \lstinline|Local_Theory.note| with signature

\begin{lstlisting}
Attrib.binding * thm list -> local_theory -> (string * thm list) * local_theory
\end{lstlisting}

Its first argument is an attribute binding (a tuple of a name and a list of attributes like \lstinline|[simp]|) paired with a list of theorems. This makes it possible to add multiple theorems at the same time, though in our case we just pass the single theorem as \lstinline|[proven_thm]|. The return values are the names that Isabelle assigned together with the list of stored theorems, and the updated local theory. Note that the name returned is always fully qualified and may therefore differ from what is passed into \lstinline|Local_Theory.note|; if referring to the fact later one should always use the returned value. Here, we do not plan to do anything further with the theorems we just declared, therefore it suffices to simply return the updated local theory.

\section{Proving from the ML Interface}\label{sec:proving}
The remaining step in the implementation of \lstinline|poly_degree| is to automatically prove the generated statement from the ML interface. This is achieved in the \lstinline|prove_correctness| function. Before we explain it, it will be helpful to first understand exactly how our proofs would look if formalised conventionally in Isabelle.

\subsection{Formalising Degree Bound Proofs}
\label{sec:proving-bounds}
Let's start with an example. We define a polynomial \lstinline|P| as well as \lstinline|P_degree| manually. The latter is exactly as it would be defined by the metaprogram in the previous \Cref{sec:declaring}. We can then state the corresponding correctness theorem and start its proof.

\isasnippet[caption={A degree bound statement set up without ML.}]{proving-example}

The first step consists of unfolding the definitions of \lstinline|P| and \mbox{\lstinline|P_degree|,} after which we obtain the goal stated in the comment. The correct lemma to apply next is the bound of \lstinline|total_degree| for the sum of polynomials. It is called \lstinline|total_degree_add|:

\begin{lstlisting}
total_degree (?P + ?Q) $\leq$ max (total_degree ?P) (total_degree ?Q)
\end{lstlisting}

However, this is not yet in a form where we can use it directly. The following statement would be more helpful, since its conclusion exactly matches our current goal.

\begin{lstlisting}
total_degree ?P $\leq$ ?d1 $\Longrightarrow$
total_degree ?Q $\leq$ ?d2 $\Longrightarrow$ total_degree (?P + ?Q) $\leq$ max ?d1 ?d2
\end{lstlisting}

We could either prove this lemma separately or derive it on the spot by combining existing inequality lemmas, as in the following method invocation.

\begin{lstlisting}
apply (rule le_trans[OF total_degree_add max.mono])
\end{lstlisting}

After applying the rule method, we obtain the following new subgoals:

\begin{lstlisting}
1. total_degree (ConstP 3 * VarP 0) $\leq$ 0 + 1
2. total_degree (VarP 0 * VarP 1) $\leq$ 1 + 1
\end{lstlisting}

To handle the new subgoals, we need to deal with the product of polynomials. This can be achieved by using \lstinline|total_degree_mult| similar to how we just applied \lstinline|total_degree_add|. After this, the remaining subgoals will only consist in proving degree bounds about the total degree of \lstinline|ConstP| or \lstinline|VarP|. The required lemmas are the following:

\begin{lstlisting}
total_degree (ConstP ?x) = 0
total_degree (VarP ?v) = 1
\end{lstlisting}

These are natural simp lemmas and have already been annotated as such. We can therefore invoke \lstinline|simp_all| to finish the proof.

To summarise, a general proof of a bound on the total degree of a polynomial consists of three steps:

\begin{enumerate}[label={\Roman*.}, ref=\Roman*]
  \item\label{item:unfold-defs} Unfold the definition of the polynomial and its degree constant.
  \item\label{item:degree-lemmas} Repeatedly apply lemmas that bound the total degree of sums, products and powers of polynomials (as long as there are still subgoals containing the sum, product or power of polynomials).
  \item\label{item:simp-all} Close the remaining goals using \lstinline|simp_all|.
\end{enumerate}

\subsection{The Degree Bound Tactic}
In Isabelle/ML, a tactic is a function that maps a single goal state to a sequence of possible next goal states, each of which would suffice to close the original goal~\cite[Section~4]{isabelle-implementation}. Every goal state is itself a list of subgoals.

The sequence of goal states is always lazily computed, which allows for backtracking as necessary when chaining multiple tactics together. If a tactic returns no next goal state, it has failed. If a tactic evaluates its next goal state to be the empty list (i.e., a goal state with no subgoals), the proof search has succeeded. Otherwise, the tactic has produced new subgoals that need to be proved.

For the Isabelle user, the Isar language provides methods such as \lstinline|rule| or \mbox{\lstinline|simp|.} Methods additionally take the proof context as input (e.g., the list of facts provided with \lstinline|using|) and are often implemented as wrappers around ML tactics. 

We implement our strategy above in a function called \lstinline|degree_correctness_tac|. It takes the context and a list of applicable theorems as input and returns an ML tactic. The three lines of the function body correspond exactly to steps I to III above.

\mlsnippet[caption={The tactic for proving degree bounds.}]{prove-correctness-tactic}

In the first step, we need to unfold the definition of the given polynomial and its degree bound. These definitional theorems will be passed as \lstinline|definitions|. The ML function corresponding to \lstinline|unfolding| in Isar is \lstinline|Local_Defs.unfold_tac|. Next to the theorems, it takes the context as an argument.

We can use the tactic combinator \lstinline|THEN| to chain multiple tactics. If the tactic to its left succeeds, then the tactic to its right is run on all resulting subgoals. This is the right behaviour for going from Step~I to Step~II as well as from Step~II to Step~III.

To implement Step~\ref{item:degree-lemmas}, we use the tactic \lstinline|resolve_tac| and pass the appropriate inequality lemmas about the degree bounds of sums, products and powers of polynomials as arguments (as well as the context). We declare the list of those lemmas separately as \lstinline|composite_degree_theorems| using \lstinline|@{thm ...}| antiquotations.

The tactic \lstinline|resolve_tac| corresponds roughly to \lstinline|apply (rule ...)| in Isar. It executes a single step of backwards reasoning on a chosen subgoal; in the simplest case, this means applying a statement $A \Longrightarrow B$ to the goal $B$, after which the new goal becomes $A$.

In our use case, we would like to run the tactic on all subgoals and not just a single one. Moreover, we want to keep applying it until it stops succeeding (which means that no sums or products of polynomials remain). This can be achieved by using the \lstinline|REPEAT| and \lstinline|SOMEGOAL| combinators. The latter applies the given tactic successively to the current subgoals until it succeeds on a subgoal and then stops.\footnote{To be precise, \lstinline|SOMEGOAL| starts at the last subgoal and works its way up to the first subgoal.} The \lstinline|REPEAT| combinator will recursively execute this behaviour until it fails.

Together, these combinators apply the given theorems exhaustively to all subgoals until none applies anymore. At this point, the subgoals should no longer contain composite polynomials, but only \lstinline|Var| and \lstinline|Const|. We can therefore proceed to Step~\ref{item:simp-all} and run the simplifier with \lstinline|simp_tac|. By default, this only applies to a specific subgoal, so we use \lstinline|ALLGOALS| to apply it to all subgoals.

\subsection{Setting Up a Proof in ML}
The final step required to run this tactic is to set up a proof state with the corresponding correctness theorem. This happens in the listing below; the function is called in the main \lstinline|poly_degree| metaprogram (\Cref{lst:poly-degree}) with the degree-bound theorem and the definitional theorems as arguments.

\mlsnippet[caption={The proof setup in ML.}]{prove-correctness-setup}

The function \lstinline|prove_correctness| initialises an Isabelle proof of the provided goal term. When it gets called in the \lstinline|poly_degree| metaprogram, the list of definitional theorems will contain \lstinline|P_def| and \lstinline|P_degree_def|; these are later passed into \lstinline|degree_correctness_tac|. The present function invokes \lstinline|Goal.prove| with the following signature.

\begin{lstlisting}
Goal.prove: Proof.context -> string list -> term list -> term ->
  ({prems: thm list, context: Proof.context} -> tactic) -> thm
\end{lstlisting}

After the first argument for the context, the second argument can be used to provide names of fixed variables (like \lstinline|fix x y| in Isar). Moreover, with the third argument, one can provide assumptions (like \lstinline|assume A|) which are passed to the tactic as \lstinline|prems|. The fourth argument is the term representing the statement to be proved, followed by the tactic, which can depend on active assumptions (\lstinline|prems|) and the context.

In our case, we do not need to set up any fixed variables or local assumptions. We simply pass the context \lstinline|lthy|, correctness theorem \lstinline|correctness_thm| and our degree bounding tactic as the appropriate arguments.

Naturally, if the tactic does not manage to close the given goal, this command will throw an error. Otherwise, it returns a value of type \lstinline|thm|, certifying that the provided term has been verified by Isabelle's kernel.

In \Cref{lst:poly-degree}, this theorem is then added to the context so that after the invocation of \lstinline|poly_degree| the correctness theorem can be used in future proofs. This completes our explanation on how to prove statements to bound polynomial degrees in Isabelle/ML.

\section{Conclusion}
In this tutorial, we introduced the basics of Isabelle/ML: computing with and constructing terms, making definitions and conducting proofs programmatically. The running example was a simplified implementation of the \lstinline|poly_degree| command, which computes upper bounds on the degrees of multivariate polynomials in Isabelle/ML. We believe that \lstinline|poly_degree| as presented here is suitable for experimentation and extension and hope that it can inspire other mathematicians to experiment with metaprogramming in Isabelle. For further reading, we refer the reader to the Isabelle implementation manual~\cite{isabelle-implementation} and the Cookbook~\cite{urban-cookbook}.

\paragraph{Outlook.} One direction for future work is to apply \lstinline|poly_degree| to the formalisation of algorithmic complexity. Here one often ends up with complicated polynomial expressions, but would prefer to state results more conceptually using big-O notation. Note that turning a polynomial bound into a big-O statement is, at heart, a degree computation; this could further be extended to support logarithms or exponential functions.

Managing the formalisation becomes more involved when bounds are nested, for example when an algorithm calls a subroutine with its own complexity bound. This is similar to the challenge of substituting polynomials into each other, which originally motivated the design of \lstinline|poly_degree|. We would be excited to see our metaprogram applied to and extended for complexity theory or further areas of mathematics.

\bibliographystyle{plainurl}
\bibliography{references}

\end{document}

%% file: isabelle_listings.tex
\usepackage{listings}
\usepackage{xcolor}
\usepackage{xparse}

\definecolor{isablue}{HTML}{006699}
\definecolor{isagreen}{HTML}{009966}
\definecolor{isared}{HTML}{BC190F}
\definecolor{isapink}{HTML}{E158D3}
\definecolor{lightblue}{HTML}{F5F9FF}

\lstdefinelanguage{isabelle}{%
  keywords=[1]{unfolding,poly_degree,poly_degree0,poly_extract,print_degree,type_synonym,datatype,fun,abbreviation,definition,proof,lemma,theorem,corollary,by,have,from,with,using,lift_definition,locale,consts,ML,theory},
  keywordstyle=[1]\bfseries\color{isablue},
  keywords=[2]{fixes,defines,where,assumes,shows,and,is,imports,begin,end,keywords},
  keywordstyle=[2]\bfseries\color{isagreen},
  keywords=[3]{if,then,else,case,of,SOME,let,in,O},
  keywordstyle=[3]\color{isablue},
  keywords=[4]{apply},
  keywordstyle=[4]\color{isared},
  commentstyle=\color{gray},
  morecomment=[l]{(*},
  morecomment=[s]{(*}{*)}
}

\lstdefinelanguage{ML}{%
  keywords=[1]{val,fun,datatype,type,structure,signature,functor,local,abstype,exception,raise,handle,fn,rec,open,infix,infixr,nonfix,op,and,andalso,orelse,withtype,sharing,include,where,eqtype},
  keywordstyle=[1]\bfseries\color{isablue},
  keywords=[2]{if,then,else,case,of,let,in,end},
  keywordstyle=[2]\bfseries\color{isagreen},
  keywords=[3]{true,false,nil,ref,NONE,SOME},
  keywordstyle=[3]\color{isablue},
  keywords=[4]{command_keyword},
  keywordstyle=[4]\itshape\bfseries,
  commentstyle=\color{isared},
  morecomment=[s]{(*}{*)},
  morestring=[b]",
  stringstyle=\color{isapink},
  showstringspaces=false,
  sensitive=true,
  backgroundcolor=\color{lightblue},
  moredelim=[s][\color{violet}]{@\{}{\}}, 
  mathescape=false
}

\NewDocumentCommand{\isasnippet}{O{} m}{%
  \ifemptyarg{#1}{%
    \lstinputlisting[language=isabelle]{isabelle_snippets/#2.txt}%
  }{%
    \lstinputlisting[language=isabelle,#1]{isabelle_snippets/#2.txt}%
  }%
}

\newcommand{\ifemptyarg}[3]{\if\relax\detokenize{#1}\relax #2\else #3\fi}

\lstdefinestyle{mllisting}{language=ML,aboveskip=0pt,frame=top,rulecolor=\color{blue!20}}

\lstnewenvironment{mllisting}[1][]
{%
  \colorbox{lightblue}{\tiny \textbf{ML}}%
  \lstset{style=mllisting,#1}%
}
{}

\NewDocumentCommand{\mlsnippet}{O{} m}{%
  \colorbox{lightblue}{\tiny \textbf{ML}}%
  \lstinputlisting[style=mllisting,#1]{isabelle_snippets/#2.txt}%
}